\documentclass[twocolumn,showpacs,prl]{revtex4}

\ifx\pdfoutput\undefined
\usepackage{graphicx}
\else
\usepackage[pdftex]{graphicx}
\usepackage{epstopdf}
\fi

\usepackage[center]{subfigure}

\begin{document}

\title{Roughness-induced critical phenomena in a turbulent flow}

\author{Nigel Goldenfeld}
\affiliation{Department of Physics, University of Illinois at
Urbana-Champaign, 1110 West Green Street, Urbana, Illinois, 61801-3080.}

\begin{abstract}

I present empirical evidence that turbulent flows are closely analogous
to critical phenomena, from a reanalysis of friction factor
measurements in rough pipes.  The data collapse found here corresponds
to Widom scaling near critical points, and implies that a full
understanding of turbulence requires explicit accounting for boundary
roughness.

\end{abstract}


\maketitle

Turbulent phenomena in fluids\cite{SREE99} are characterized by strong
fluctuations and power-law spectra\cite{FRIS95} that are
suggestive\cite{EYIN94} of the power-law correlations observed in
critical phenomena at continuous phase transitions\cite{NGBook}.
Attempts to link these two sets of phenomena have primarily, but not
exclusively\cite{CHOR97,BRAM98}, focused on the calculation of scaling
exponents\cite{SREE97} in ideal systems which are isotropic and
homogeneous, neglecting boundaries. However, no analogue has been found
for the wide variety of thermodynamic scaling phenomena\cite{WIDO65}
that are an equally fundamental aspect of continuous phase transitions,
and whose elucidation\cite{KADA66} led to a complete understanding of
critical phenomena\cite{WILS71}.  Here we consider the important role
of boundary roughness on fluid flow, by reanalyzing Nikuradse's
experimental measurements\cite{NIKU33} of the friction exerted on a
turbulent fluid by the walls of a rough pipe.  We show that the data as
a function of Reynolds number and relative roughness collapse onto one
universal curve, when appropriately scaled. This analogue of Widom
scaling\cite{WIDO65} implies that boundary roughness must be included
in a complete description of turbulence, and establishes the long
sought-after precise connection to critical phenomena.

Turbulent flows are characterised by their Reynolds number, defined as
$Re\equiv UL/\nu$, where $U$ is a typical velocity at the length scale
$L$, and $\nu$, the kinematic viscosity, is the viscosity of the fluid
divided by its density. In 1941, Kolmogorov\cite{KOLM41} and
Obukhov\cite{OBUK41}, recognised that at large enough Reynolds numbers,
fluid motion is, over a wide range of length scales, a dynamical,
energy-conserving but irregular swirling motion\cite{SREE99} governed
by inertia, rather than a dissipative phenomenon. Thus, they pointed
out that for this so-called inertial range of scales, observables
should be independent of $\nu$. In particular, for the inertial range,
the turbulent energy spectrum $E$ of longitudinal velocity fluctuations
$\delta v_k$ in wavenumber space, $k$, can only depend upon the mean
energy transfer rate $\bar\epsilon$ and $k$ itself in a manner dictated
by dimensional analysis: $E(k)\equiv \left<|\delta
v_k|^2\right>=\bar\epsilon k^{-5/3}$. This
experimentally-verified\cite{GRAN62} power-law scaling (often referred
to as K41) applies on small scales in a turbulent flow, but not so
small that molecular viscosity becomes important. The existence of a
wide range of length scales, over which power-law (and thus
scale-invariant) correlated fluctuations are found, is
reminiscent\cite{EYIN94} of the power-law fluctuations on many length
scales that accompany critical phenomena\cite{NGBook}: for example, in
a ferromagnet near its critical point, the Fourier component of the
magnetisation $M$ at wavenumber $k$ satisfies $G(k)\equiv
\left<|M_k|^2\right> \propto k^{-2+\eta}$, where $\eta$ is an anomalous
scaling exponent that describes departures from mean field theory. This
power-law scaling applies at large scales, and is independent of the
small scale details of the system, such as the nature of the crystal
lattice.

Power-law scaling of correlation functions is, however, only one of two
key aspects of critical phenomena\cite{NGBook}.  The other, equally
important aspect is the phenomenon of data collapse, or Widom
scaling\cite{WIDO65}: for example, in a ferromagnet, the equation of
state, nominally a function of two variables, is expressible in terms
of a single reduced variable that depends on a combination of external
field and temperature.  What is the turbulent analogue to Widom scaling
in thermodynamics?  To address this question, it is necessary to
examine data on the {\it large-scale} properties of turbulence, for
example the friction factor in pipe flow.

\begin{figure}
\includegraphics[width=\columnwidth]{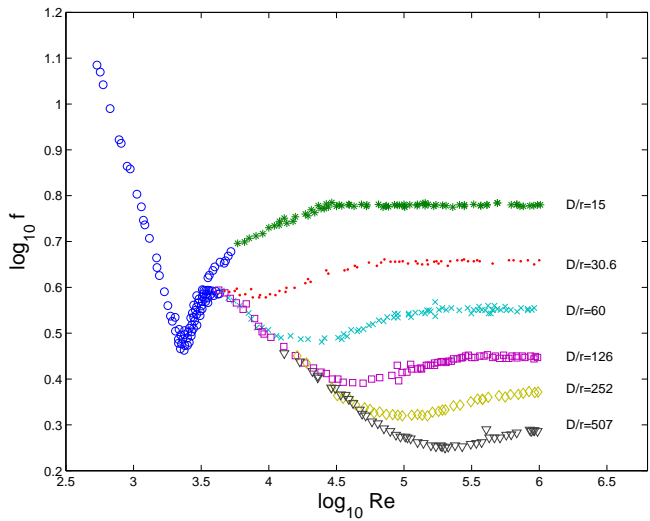}
\caption{(Color online) Friction factor for turbulent flow in a rough pipe, as
reported by Nikuradse\cite{NIKU33}.  The data were measured at
different values of $Re$ and $r/D$, and extracted from Nikuradse's
tabular and graphical presentation.}\label{Nikuradse-data-raw}
\end{figure}

In 1932 and 1933, Nikuradse undertook a seminal series of measurements
of flow in nominally smooth\cite{NIKU32} and rough pipes\cite{NIKU33},
measuring {\it inter alia\/} the friction factor $f$, related to the
pressure drop across the pipe\cite{JIME04, MCKE05}.  These measurements
have remained the benchmark in the field, being the only systematic
measurements of a single flow geometry over such a wide range of
Reynolds numbers. Nikuradse's experiments on sand-roughened pipes used
sand grains of a well-defined size $r$, repeated over a wide range of
values of $r$ and pipes of different radii $D$.  Nikuradse was able to
verify the expectation of hydrodynamic similarity: the flow properties
depend on the roughness only through the combination $r/D$. Nikuradse
presented results for the shear force per unit area $\tau$ exerted by
the flow on the walls of the pipe in the form $f=\tau/\rho U^2$, and
these data are plotted in figure (\ref{Nikuradse-data-raw}).


There are several broad features to note in these data.  First, in the
smoothest pipes studied, for $Re < 100,000$, in the turbulent regime,
the friction factor is a decreasing function of $Re$, varying to a good
approximation in a manner usually attributed to Blasius\cite{BLAS13,
SCHL79} as $f\sim Re^{-1/4}$.  As the roughness scale is increased, and
thus for smaller values of $D/r$, the extent over which the Blasius
scaling extends becomes smaller.  Thus, we can represent this feature
by the statement that $f\sim Re^{-1/4}$ asymptotically as $ r/D
\rightarrow 0$. Asymptotically for rough pipes at large $Re$, the
friction becomes {\it independent\/} of $Re$, and depends only on the
roughness $r/D$, varying to a good approximation, according to
Strickler's law\cite{STRI23}, as $f \sim (r/D)^{1/3}$.  These broad
characteristics are also visible in aggregate in other pipe flow
data\cite{BAUE36,GALA39,COLE37,COLE39,SLET03}, but no other single data
set captures the full range shown here, with as little scatter evident
in the data.

These features place strong constraints on the functional form of the
friction factor $f(Re, r/D)$.  In fact, these constraints precisely
parallel those on thermodynamic properties of ferromagnets near the
critical temperature $T_c$, as a function of reduced temperature
$t=|T-T_c|/T_c$ and external magnetic field $H$. For example, at the
critical temperature, the magnetic equation of state has the form $M
\sim H^{1/\delta}$ for $t=0$, where $\delta$ is a critical exponent
whose value can be computed by renormalisation group (RG)
theory\cite{NGBook}; and for zero field, the magnetisation continuously
approaches zero as $T\rightarrow {T_c}^-$ with a power law variation
$M\sim t^\beta$ for $H=0$, where $\beta$ is another critical exponent
whose value can be computed by RG.  Widom\cite{WIDO65} discovered that
these properties all followed if the thermodynamic free energy obeyed
certain scaling properties, later shown to follow from renormalization
group considerations\cite{KADA66, WILS71}.

We connect scaling in turbulence with that in critical phenomena by
observing that the limit $Re\rightarrow \infty$ is analogous to the
limit $t^{-1}\rightarrow\infty$, whereas the limit $H\rightarrow 0$ is
analogous to the limit $ r/D \rightarrow 0$.  Thus, the analytic
properties of the friction factor can be derived if we follow Widom's
scaling argument\cite{WIDO65}, and propose a scaling form for the
friction factor: $f(Re, r/D)= Re^{-1/4} g(Re^{\alpha} r/D)$, where
$g(z)$ is an unknown scaling function of a single variable $z$, which
tends to a constant for small values of $z$, and $\alpha$ is an
exponent that we can determine by the requirement that the $Re$
dependence should cancel out of the formula for $f$ at large $Re$,
leaving the Strickler law $f \sim (r/D)^{1/3}$. This requires that
$g(z) \sim z^{1/3}$ as $z\rightarrow \infty$, and therefore
$\alpha/3=1/4$. Thus, we conclude that
\begin{equation}
f=Re^{-1/4} g(Re^{3/4} (r/D)).
\label{eqn:scaling}
\end{equation}

\begin{figure}
\includegraphics[width=\columnwidth]{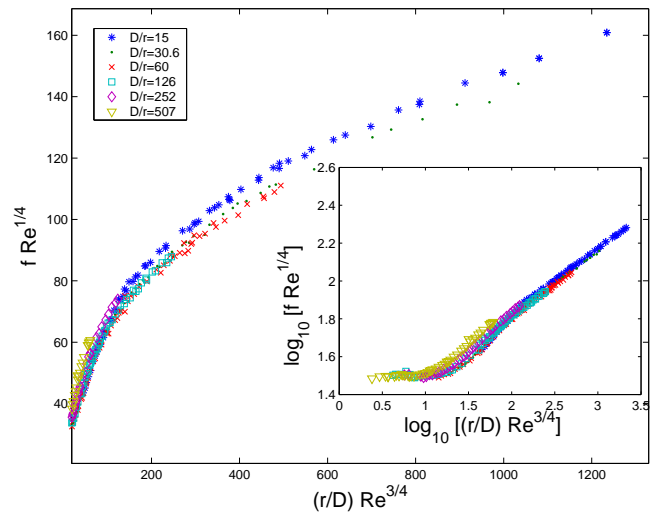}
\caption{(Color online) Friction factor for turbulent flow in a rough
pipe, as reported by Nikuradse\cite{NIKU33}, scaled according to the
text. Inset: the same plot on a logarithmic scale.}
\label{Nikuradse-data-scaled}
\end{figure}

The scaling form of Eq. (\ref{eqn:scaling}) predicts that the turbulent
friction factor data, measured as a function of both $r/D$ and $Re$,
and thus in principle occupying a two dimensional space, will actually
collapse onto a one-dimensional curve, when plotted as $f Re^{1/4}$
versus $(r/D) Re^{3/4}$.  The test of this prediction is shown in
figure (\ref{Nikuradse-data-scaled}), where Nikuradse's data occupying
the plane of figure (\ref{Nikuradse-data-raw}) collapses onto a single
curve when plotted in the reduced variables of Eq. (\ref{eqn:scaling}).
Note that the data collapse occurs for those data that lie between the
Blasius and Strickler regimes only.  Small deviations from the data
collapse are visible, but it is not clear to what extent these reflect
uncertainties in the data\cite{WILL51} or something more fundamental.

The scaling function that we have extracted from the data is unlikely
to be universal in the sense of being independent of the nature of the
roughness of the pipe.  By analogy with the effects of long-range
interactions in critical spin systems\cite{NGBook}, it seems probable
that self-affine roughness\cite{SLET03} will have a different effect on
the flow than periodic single-scale roughness, and this can be
reflected in the scaling function, the scaling exponents, or both.
Inspection of the data from different pipe flow experiments suggests
that the first possibility is the most likely, but this remains to be
checked in detail.

The interpretation of the analogy discovered here is very natural.  In
magnetic systems, for example, the power-law scaling of $G(k)$ is now
understood to be associated with the fact that magnetisation can be
induced in a magnet by the application of an external magnetic field
$H$.  The sensitivity of the order in a magnet to a perturbation by $H$
becomes exquisite near a critical point, and thermodynamic variables
contain a singular dependence on $H$ and $t$.  The fact that the
boundary roughness plays the role of external magnetic field is a
reflection of the fact that small perturbations couple to the flow and
drive the turbulent state.

The exponents in the scaling theory given here are taken from
experiment; however, Gioia and Chakraborty have recently pointed
out\cite{GIOI05} that the friction factor $f$ can be related to the
local structure of turbulence, by considering the momentum flux at the
pipe boundary.  The Nikuradse data show four features: a hump, the
Blasius regime, a shallow minimum, and the Strickler regime.  The
scaling argument presented here implies that the Blasius and Strickler
regimes are both manifestations of inertial range scaling coupled with
wall friction, and indeed, Gioia and Chakraborty find from momentum
flux considerations that this is sufficient to reproduce the Blasius
and Strickler regimes.  If they then include the dissipation range in
their formula for the friction factor, they find that this reproduces
the shallow minimum between the Blasius and Strickler regime.
Interesting their formulae for the pure inertial and the inertial +
dissipation ranges both satisfy Eq. (\ref{eqn:scaling}).  The hump
in the friction factor arises from the energy-containing range, lies
outside the range bracketed by Blasius and Strickler and is absent from
their predictions if this spectral feature is not included in their
formula. Thus, in summary, the features which are described by Eq.
(\ref{eqn:scaling}) are indeed related to the spectral features of the
inertial range (and the dissipation range).  Thus, just as in critical
phenomena, the large-scale phenomenology of turbulence can be related
to the power law fluctuations.  Presumably, the scaling result Eq.
(\ref{eqn:scaling}) can be derived from a renormalization group
argument\cite{KADA66}.

Eq. (\ref{eqn:scaling}) describes the nonequilibrium driven steady
state of fully-developed turbulence, which is to be contrasted with the
phenomenon of the transition from laminar to turbulent flow.  In the
case of turbulent pipe flow, it is generally accepted that laminar
Hagen-Poiseuille flow is linearly stable at all Reynolds
numbers, and that there is a finite amplitude instability
to turbulence\cite{PFEN61, HOF04}, whose amplitude varies\cite{HOF03}
as $Re^{-1}$.  Thus, this transition has the character of a first order
transition in thermodynamics, and is not related to the theory given
here.

Our results establish boundary roughness as a key element in a proper
theoretical description of turbulence, in the same way that a proper
understanding of the ferromagnetic critical point at zero field would
not be possible without taking into account the behaviour for non-zero
values of $H$.   Our analysis highlights the need for a definitive set
of experiments to replicate Nikuradse's data set, with a view to
greater precision and removing sources of uncertainty\cite{WILL51} in
both the data set and in the widely-used Colebrook\cite{COLE39}
semi-empirical fit\cite{SMIT05}. In ongoing work we are exploring the
effects of roughness in two-dimensional turbulence\cite{KELL02}, where
predictions for the analogues of the Blasius and Strickler laws reveal
interesting differences from the three dimensional results.

I acknowledge valuable discussions with G.I. Barenblatt, Gustavo Gioia
and Pinaki Chakraborty, whom I also thank for assistance with the
figures.  This work was supported in part by the National Science
Foundation through grant number NSF-EAR-02-21743.

\bibliographystyle{apsrev}

\bibliography{turbulence}

\begin{thebibliography}{32}
\expandafter\ifx\csname natexlab\endcsname\relax\def\natexlab#1{#1}\fi
\expandafter\ifx\csname bibnamefont\endcsname\relax
  \def\bibnamefont#1{#1}\fi
\expandafter\ifx\csname bibfnamefont\endcsname\relax
  \def\bibfnamefont#1{#1}\fi
\expandafter\ifx\csname citenamefont\endcsname\relax
  \def\citenamefont#1{#1}\fi
\expandafter\ifx\csname url\endcsname\relax
  \def\url#1{\texttt{#1}}\fi
\expandafter\ifx\csname urlprefix\endcsname\relax\def\urlprefix{URL }\fi
\providecommand{\bibinfo}[2]{#2}
\providecommand{\eprint}[2][]{\url{#2}}

\bibitem[{\citenamefont{Sreenivasan}(1999)}]{SREE99}
\bibinfo{author}{\bibfnamefont{K.~R.} \bibnamefont{Sreenivasan}},
  \bibinfo{journal}{Rev. Mod. Phys.} \textbf{\bibinfo{volume}{71}},
  \bibinfo{pages}{S383} (\bibinfo{year}{1999}).

\bibitem[{\citenamefont{Frisch}(1995)}]{FRIS95}
\bibinfo{author}{\bibfnamefont{U.}~\bibnamefont{Frisch}},
  \emph{\bibinfo{title}{Turbulence: The Legacy of A.N. Kolmogorov}}
  (\bibinfo{publisher}{Cambridge University Press}, \bibinfo{year}{1995}).

\bibitem[{\citenamefont{Eyink and Goldenfeld}(1994)}]{EYIN94}
\bibinfo{author}{\bibfnamefont{G.}~\bibnamefont{Eyink}} \bibnamefont{and}
  \bibinfo{author}{\bibfnamefont{N.}~\bibnamefont{Goldenfeld}},
  \bibinfo{journal}{Phys. Rev. E} \textbf{\bibinfo{volume}{50}},
  \bibinfo{pages}{4679} (\bibinfo{year}{1994}).

\bibitem[{\citenamefont{Goldenfeld}(1992)}]{NGBook}
\bibinfo{author}{\bibfnamefont{N.}~\bibnamefont{Goldenfeld}},
  \emph{\bibinfo{title}{Lectures on phase transitions and the renormalization
  group}} (\bibinfo{publisher}{Addison-Wesley}, \bibinfo{year}{1992}).

\bibitem[{\citenamefont{Chorin}(1997)}]{CHOR97}
\bibinfo{author}{\bibfnamefont{A.~J.} \bibnamefont{Chorin}},
  \emph{\bibinfo{title}{Vorticity and turbulence}}
  (\bibinfo{publisher}{Springer}, \bibinfo{year}{1997}).

\bibitem[{\citenamefont{Bramwell et~al.}(1998)\citenamefont{Bramwell,
  Holdsworth, and Pinton}}]{BRAM98}
\bibinfo{author}{\bibfnamefont{S.~T.} \bibnamefont{Bramwell}},
  \bibinfo{author}{\bibfnamefont{P.~C.~W.} \bibnamefont{Holdsworth}},
  \bibnamefont{and} \bibinfo{author}{\bibfnamefont{J.~F.}
  \bibnamefont{Pinton}}, \bibinfo{journal}{Nature}
  \textbf{\bibinfo{volume}{396}}, \bibinfo{pages}{552} (\bibinfo{year}{1998}).

\bibitem[{\citenamefont{Sreenivasan and Antonia}(1997)}]{SREE97}
\bibinfo{author}{\bibfnamefont{K.~R.} \bibnamefont{Sreenivasan}}
  \bibnamefont{and} \bibinfo{author}{\bibfnamefont{R.~A.}
  \bibnamefont{Antonia}}, \bibinfo{journal}{Annu. Rev. Fluid Mech.}
  \textbf{\bibinfo{volume}{29}}, \bibinfo{pages}{435} (\bibinfo{year}{1997}).

\bibitem[{\citenamefont{Widom}(1965)}]{WIDO65}
\bibinfo{author}{\bibfnamefont{B.}~\bibnamefont{Widom}}, \bibinfo{journal}{J.
  Chem. Phys.} \textbf{\bibinfo{volume}{43}}, \bibinfo{pages}{3898}
  (\bibinfo{year}{1965}).

\bibitem[{\citenamefont{Kadanoff}(1966)}]{KADA66}
\bibinfo{author}{\bibfnamefont{L.~P.} \bibnamefont{Kadanoff}},
  \bibinfo{journal}{Physics} \textbf{\bibinfo{volume}{2}}, \bibinfo{pages}{263}
  (\bibinfo{year}{1966}).

\bibitem[{\citenamefont{Wilson}(1971)}]{WILS71}
\bibinfo{author}{\bibfnamefont{K.~G.} \bibnamefont{Wilson}},
  \bibinfo{journal}{Phys. Rev. B} \textbf{\bibinfo{volume}{4}},
  \bibinfo{pages}{3174} (\bibinfo{year}{1971}).

\bibitem[{\citenamefont{Nikuradze}(1933)}]{NIKU33}
\bibinfo{author}{\bibfnamefont{J.}~\bibnamefont{Nikuradze}}
  (\bibinfo{year}{1933}), \bibinfo{note}{vDI Forschungsheft, vol. 361 [In
  English, in Technical Memorandum 1292, National Advisory Committee for
  Aeronautics (1950).]}.

\bibitem[{\citenamefont{Kolmogorov}(1941)}]{KOLM41}
\bibinfo{author}{\bibfnamefont{A.~N.} \bibnamefont{Kolmogorov}},
  \bibinfo{journal}{Dokl. Akad. Nauk. SSSR} \textbf{\bibinfo{volume}{30}},
  \bibinfo{pages}{299} (\bibinfo{year}{1941}), \bibinfo{note}{[English
  translation in Proc. R. Soc. London Ser. A 434 (1991)]}.

\bibitem[{\citenamefont{Obukhov}(1941)}]{OBUK41}
\bibinfo{author}{\bibfnamefont{A.~M.} \bibnamefont{Obukhov}},
  \bibinfo{journal}{Dokl. Akad. Nauk. SSSR} \textbf{\bibinfo{volume}{32}},
  \bibinfo{pages}{22} (\bibinfo{year}{1941}).

\bibitem[{\citenamefont{Grant et~al.}(1962)\citenamefont{Grant, Stewart, and
  Moilliet}}]{GRAN62}
\bibinfo{author}{\bibfnamefont{H.~L.} \bibnamefont{Grant}},
  \bibinfo{author}{\bibfnamefont{R.~W.} \bibnamefont{Stewart}},
  \bibnamefont{and} \bibinfo{author}{\bibfnamefont{A.}~\bibnamefont{Moilliet}},
  \bibinfo{journal}{J. Fluid Mech.} \textbf{\bibinfo{volume}{12}},
  \bibinfo{pages}{241} (\bibinfo{year}{1962}).

\bibitem[{\citenamefont{Nikuradze}(1932)}]{NIKU32}
\bibinfo{author}{\bibfnamefont{J.}~\bibnamefont{Nikuradze}}
  (\bibinfo{year}{1932}), \bibinfo{note}{vDI Forschungsheft, vol. 356 [In
  English, in NASA TT F-10, 359 (1966).]}.

\bibitem[{\citenamefont{Jim\'enez}(2004)}]{JIME04}
\bibinfo{author}{\bibfnamefont{J.}~\bibnamefont{Jim\'enez}},
  \bibinfo{journal}{Ann. Rev. Fluid Mech.} \textbf{\bibinfo{volume}{36}},
  \bibinfo{pages}{173} (\bibinfo{year}{2004}).

\bibitem[{\citenamefont{McKeon et~al.}(2005)\citenamefont{McKeon, Zagarola, and
  Smits}}]{MCKE05}
\bibinfo{author}{\bibfnamefont{B.~J.} \bibnamefont{McKeon}},
  \bibinfo{author}{\bibfnamefont{M.~V.} \bibnamefont{Zagarola}},
  \bibnamefont{and} \bibinfo{author}{\bibfnamefont{A.~J.} \bibnamefont{Smits}},
  \bibinfo{journal}{J. Fluid Mech.} \textbf{\bibinfo{volume}{538}},
  \bibinfo{pages}{429} (\bibinfo{year}{2005}).

\bibitem[{\citenamefont{Blasius}(1913)}]{BLAS13}
\bibinfo{author}{\bibfnamefont{H.}~\bibnamefont{Blasius}}
  (\bibinfo{year}{1913}), \bibinfo{note}{forsch. Arb. Ing. Wes. No. 134,
  Berlin}.

\bibitem[{\citenamefont{Schlichting}(1979)}]{SCHL79}
\bibinfo{author}{\bibfnamefont{H.}~\bibnamefont{Schlichting}},
  \emph{\bibinfo{title}{Boundary layer theory}}
  (\bibinfo{publisher}{McGraw-Hill}, \bibinfo{year}{1979}).

\bibitem[{\citenamefont{Strickler}(1923)}]{STRI23}
\bibinfo{author}{\bibfnamefont{A.}~\bibnamefont{Strickler}}
  (\bibinfo{year}{1923}), \bibinfo{note}{mitteilungen des Eidgenossischen Amtes
  fur Wasserwirtschaft 16, Bern, Switzerland ~Translated as \lq\lq
  Contributions to the question of a velocity formula and roughness data for
  streams, channels and closed pipelines." by T. Roesgan and W. R. Brownie,
  Translation T-10, W. M. Keck Lab of Hydraulics and Water Resources, Calif.
  Inst. Tech., Pasadena, Calif. January 1981}.

\bibitem[{\citenamefont{Bauer and Galavics}(1936)}]{BAUE36}
\bibinfo{author}{\bibfnamefont{B.}~\bibnamefont{Bauer}} \bibnamefont{and}
  \bibinfo{author}{\bibfnamefont{F.}~\bibnamefont{Galavics}},
  \bibinfo{journal}{Archiv f\"ur Waermewirtschaft}
  \textbf{\bibinfo{volume}{17}}, \bibinfo{pages}{125} (\bibinfo{year}{1936}).

\bibitem[{\citenamefont{Galavics}(1939)}]{GALA39}
\bibinfo{author}{\bibfnamefont{F.}~\bibnamefont{Galavics}},
  \bibinfo{journal}{Schweizer Archiv} \textbf{\bibinfo{volume}{5}},
  \bibinfo{pages}{337} (\bibinfo{year}{1939}).

\bibitem[{\citenamefont{Colebrook and White}(1937)}]{COLE37}
\bibinfo{author}{\bibfnamefont{C.~F.} \bibnamefont{Colebrook}}
  \bibnamefont{and} \bibinfo{author}{\bibfnamefont{C.~M.} \bibnamefont{White}},
  \bibinfo{journal}{Proc. Roy. Soc. Lon, Ser. A}
  \textbf{\bibinfo{volume}{161}}, \bibinfo{pages}{367} (\bibinfo{year}{1937}).

\bibitem[{\citenamefont{Colebrook}(1939)}]{COLE39}
\bibinfo{author}{\bibfnamefont{C.~F.} \bibnamefont{Colebrook}},
  \bibinfo{journal}{Institution of Civ. Eng. Journal}
  \textbf{\bibinfo{volume}{11}}, \bibinfo{pages}{133} (\bibinfo{year}{1939}).

\bibitem[{\citenamefont{Sletfjerding and Gudmundsson}(2003)}]{SLET03}
\bibinfo{author}{\bibfnamefont{E.}~\bibnamefont{Sletfjerding}}
  \bibnamefont{and} \bibinfo{author}{\bibfnamefont{J.~S.}
  \bibnamefont{Gudmundsson}}, \bibinfo{journal}{J. Energy Res. Tech.}
  \textbf{\bibinfo{volume}{125}}, \bibinfo{pages}{126} (\bibinfo{year}{2003}).

\bibitem[{\citenamefont{Williamson}(1951)}]{WILL51}
\bibinfo{author}{\bibfnamefont{J.}~\bibnamefont{Williamson}},
  \bibinfo{journal}{La Houille Blanche} \textbf{\bibinfo{volume}{6}},
  \bibinfo{pages}{738} (\bibinfo{year}{1951}).

\bibitem[{\citenamefont{Gioia and Chakraborty}(2005)}]{GIOI05}
\bibinfo{author}{\bibfnamefont{G.}~\bibnamefont{Gioia}} \bibnamefont{and}
  \bibinfo{author}{\bibfnamefont{P.}~\bibnamefont{Chakraborty}}
  (\bibinfo{year}{2005}), \bibinfo{note}{submitted to Physical Review Letters.
  Available at http://arxiv.org/abs/physics/0507066}.

\bibitem[{\citenamefont{Pfenniger}(1961)}]{PFEN61}
\bibinfo{author}{\bibfnamefont{W.}~\bibnamefont{Pfenniger}}, in
  \emph{\bibinfo{booktitle}{Boundary Layer and Flow Control}}, edited by
  \bibinfo{editor}{\bibfnamefont{G.~V.} \bibnamefont{Lachman}}
  (\bibinfo{publisher}{Pergamon, New York}, \bibinfo{year}{1961}), pp.
  \bibinfo{pages}{961--980}.

\bibitem[{\citenamefont{Hof et~al.}(2004)}]{HOF04}
\bibinfo{author}{\bibfnamefont{B.}~\bibnamefont{Hof}} \bibnamefont{et~al.},
  \bibinfo{journal}{Science} \textbf{\bibinfo{volume}{305}},
  \bibinfo{pages}{1594} (\bibinfo{year}{2004}).

\bibitem[{\citenamefont{Hof et~al.}(2003)\citenamefont{Hof, Juel, and
  Mullin}}]{HOF03}
\bibinfo{author}{\bibfnamefont{B.}~\bibnamefont{Hof}},
  \bibinfo{author}{\bibfnamefont{A.}~\bibnamefont{Juel}}, \bibnamefont{and}
  \bibinfo{author}{\bibfnamefont{T.}~\bibnamefont{Mullin}},
  \bibinfo{journal}{Phys. Rev. Lett.} \textbf{\bibinfo{volume}{91}},
  \bibinfo{pages}{2445021} (\bibinfo{year}{2003}).

\bibitem[{\citenamefont{Smits et~al.}(2005)\citenamefont{Smits, Shockling, and
  Allen}}]{SMIT05}
\bibinfo{author}{\bibfnamefont{A.~J.} \bibnamefont{Smits}},
  \bibinfo{author}{\bibfnamefont{M.}~\bibnamefont{Shockling}},
  \bibnamefont{and} \bibinfo{author}{\bibfnamefont{J.}~\bibnamefont{Allen}}, in
  \emph{\bibinfo{booktitle}{4th AIAA Theoretical Fluid Mechanics Meeting}}
  (\bibinfo{year}{2005}).

\bibitem[{\citenamefont{Kellay and Goldburg}(2002)}]{KELL02}
\bibinfo{author}{\bibfnamefont{H.}~\bibnamefont{Kellay}} \bibnamefont{and}
  \bibinfo{author}{\bibfnamefont{W.~I.} \bibnamefont{Goldburg}},
  \bibinfo{journal}{Rep. Prog. Phys.} \textbf{\bibinfo{volume}{65}},
  \bibinfo{pages}{845} (\bibinfo{year}{2002}).

\end{thebibliography}

\end{document}